# Interlayer Coupling and Strain Localization in Small-Twist-Angle Graphene Flakes

Vahid Morovati,[a] Zhiming Xue,[b] Kenneth M. Liechti,[a] Rui Huang[a]

[a] *Department of Aerospace Engineering and Engineering Mechanics, University of Texas, Austin, TX 78712, USA*
[b] *Center for Composite Materials, Harbin Institute of Technology, Harbin 150001, China*

## Abstract

Twisted bilayer graphene (TBG) exhibits a wide range of intriguing physical properties, such as superconductivity, ferromagnetism, and superlubricity. Depending on the twist angle, periodic moiré superlattices form in twisted bilayer graphene, with inhomogeneous interlayer coupling and lattice deformation. For a small twist angle (typically < 2°), each moiré supercell contains a large number of atoms (>10,000), making it computationally expensive for first-principles and atomistic modeling. In this work, a finite element method based on a continuum model is used to simulate the inhomogeneous interlayer and intralayer deformations of twisted graphene flakes on a rigid graphene substrate. The van der Waals interactions between the graphene layers are described by a periodic potential energy function, whereas the graphene flake is treated as a continuum membrane with effective elastic properties. Our simulations show that structural relaxation and the induced strain localization are most significant in a relatively large graphene flake at small twist angles, where the strain distribution is highly localized as shear strain solitons along the boundaries between neighboring domains of commensurate AB stacking. Moreover, it is found that there exist many metastable equilibrium configurations at particular twist angles, depending on the flake size. The nonlinear mechanics of twisted bilayer graphene is thus expected to be essential for understanding the strain distributions in the moiré superlattices and the strain effects on other physical properties.

# 1. Introduction

The successful isolation of monolayer graphene by micromechanical exfoliation of bulk graphite opened up the research field of two-dimensional (2D) materials (Novoselov et al., 2005). More recently, a new paradigm for material design has emerged by stacking 2D materials on top of one another (Geim and Grigorieva, 2013; Novoselov et al., 2016). The resulting multilayered structures, often called van der Waals (vdW) materials or heterostructures, feature strong intralayer covalent bonds and relatively weak interlayer vdW interactions. These highly anisotropic interactions provide the vdW materials with tunable collective properties via the stacking sequence and the relative twisting or straining between the atomic layers (Kim et al., 2016 & 2017; Cao et al., 2018; Frisenda et al., 2018). Such an unprecedented level of tunability makes the vdW materials attractive for a wide range of technological applications (e.g., photodetectors, photovoltaics and light-emitting devices) (Liu et al., 2016). In order to fulfill the promising applications of the multilayered vdW materials and heterostructures, it is critically important to understand the mechanics at the interfaces between various 2D materials (Dai et al., 2020), where the interlayer mechanical interactions such as adhesion/separation and friction are coupled intimately with the intralayer lattice deformations of the 2D materials. In particular, a number of intriguing phenomena have been observed in twisted bilayer graphene, such as strain solitons (Alden et al., 2013) and 2D moiré superlattices (Yoo et al., 2019; McGilly et al., 2020), along with extraordinary physical properties, such as unconventional superconductivity (Cao et al., 2018), emergent ferromagnetism (Sharpe et al., 2019), and superlubricity (Feng et al., 2013).

For bilayer graphene, the commensurate Bernal stacking (i.e., AB and BA) is most stable with the minimum interlayer potential energy. When two graphene layers are not stacked perfectly in the commensurate state, such as twisted bilayer graphene, the interlayer potential energy is higher and can be reduced by forming periodic domains with alternating AB and BA stacking, separated by strain solitons (or domain walls). As a result, a periodic 2D moiré superlattice is formed with a period depending on the relative twist angle (Bistritzer and MacDonald, 2011; Hermann, 2012). For a small twist angle (typically < 2°), the moiré period ($\lambda_M$) ranges from tens to hundreds of nanometers. Although the characteristic moiré length scale $\lambda_M$ has been well established by assuming the graphene lattices to be rigid, it remains unclear how the mechanics of the highly deformable graphene with interlayer coupling could lead to commensurate-incommensurate phase transitions and various domain structures (Popov et al., 2011; Dai et al., 2016a; Jain et al., 2017; Zhang and Tadmor, 2018; Guinea and Walet, 2019).

A variety of modeling and simulation approaches have been developed to understand the mechanics of 2D moiré superlattices. Fully atomistic simulations (DFT and MD) are typically limited to periodic moiré supercells of relatively small sizes (van Wijk, et al., 2015; Guinea and Walet, 2019; Zhu et al., 2019; Bagchi, et al., 2020). To overcome this limitation, a multiscale approach was proposed by combining DFT calculations of the interlayer potential function (also called generalized stacking fault energy or GSFE) with a continuum elastic plate model (Kumar et al., 2015 and 2016; Dai et al., 2016a). Similarly, a discrete-continuum model was developed by including full atomistic interactions in the short range (using empirical interatomic potentials) along with a continuum model for long-range interactions (Zhang and Tadmor, 2017 and 2018). Alternatively, a continuum dislocation model of 2D moiré superlattices was proposed, in which the strain solitons (or domain walls) are treated as interlayer dislocations or van der Waals dislocations (Butz et al., 2014; Dai et al., 2016b; Pochet et al., 2017; Annevelink et al., 2020). While these modeling and simulations have provided significant insights into the mechanics of structural relaxation in 2D moiré superlattices, the computational cost remains high even with the multiscale approach, which limits the model size (typically < 100 nm) and thus cannot simulate 2D moiré superlattices at small twist angles. In this work, we present a finite element method based on a continuum model to simulate the interlayer and intralayer deformations of twisted graphene flakes on a rigid graphene substrate. The continuum model is atomistically informed and complements the first-principles based atomistic models with the potential of scaling up for larger systems.

The remainder of this paper is organized as follows. Section 2 describes the continuum model with a periodic interlayer potential function for the van der Waals interactions between graphene layers. In Section 3, we consider twisting of a rigid graphene flake, without intralayer lattice deformation of the flake. Section 4 presents numerical results for twisted elastic graphene flakes, and discusses the strain distributions and stability. We conclude in Section 5 with a summary of the findings.

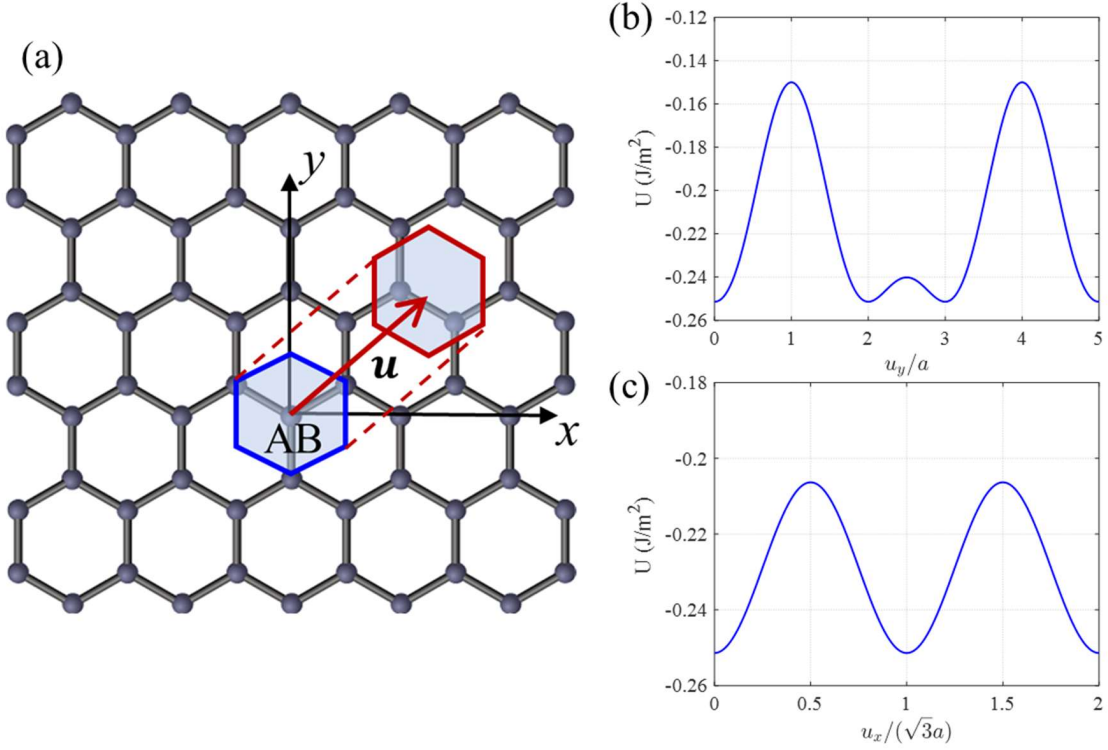

**Figure 1**. (a) Schematic illustration of a hexagonal cell of graphene on top of another graphene monolayer, with a relative in-plane displacement $\boldsymbol{u}$ from a commensurate AB stacking (blue) to an incommensurate stacking (red). (b-c) Periodic corrugations of the interlayer potential energy with respect to the relative in-plane displacements in the armchair ($u_y$) and zigzag ($u_x$) directions. The interlayer separation is fixed at $z_0 = 0.334$ nm ($u_z = 0$) for this calculation.

## 2. An interlayer potential function for van der Waals interactions

It has been shown that the interaction potential energy between two graphene layers can be written as a function of the relative displacements (Kumar et al., 2015 and 2016; Xue et al., 2022), namely

$$U(u_x, u_y, u_z) = U_0(u_z) + U_1(u_z) f(u_x, u_y), \qquad (1)$$

where $u_x$ and $u_y$ are relative in-plane displacements, and $u_z$ is the relative normal displacement. Start from an equilibrium AB stacking with $u_x = u_y = u_z = 0$ (see Fig. 1a). The first term $U_0(u_z)$ describes the dependence of the interaction potential energy on the normal separation ($u_z$) in the commensurate AB stacking ($u_x = u_y = 0$), and the second term describes the periodic corrugation of the potential energy with respect to the in-plane displacements through a function, $f(u_x, u_y)$, with a corrugation amplitude depending on the normal separation through $U_1(u_z)$. Following a previous work (Xue et. al 2022), we write

$$U_0(u_z) = \Gamma_0 \left[ -\frac{5}{3}\left(1 + \frac{u_z}{z_0}\right)^{-4} + \frac{2}{3}\left(1 + \frac{u_z}{z_0}\right)^{-10} \right], \quad (2)$$

$$U_1(u_z) = \eta\Gamma_0 \left[ -\left(1 + \frac{u_z}{z_0}\right)^{-4} + \beta\left(1 + \frac{u_z}{z_0}\right)^{-10} \right], \quad (3)$$

$$f(u_x, u_y) = \frac{3}{2} + \cos\left(G_1\left(\frac{u_y}{a} - 1\right)\right) + 2\cos\left(\frac{G_1}{2}\left(\frac{u_y}{a} - 1\right)\right)\cos\left(\frac{\sqrt{3}G_1}{2}\frac{u_x}{a}\right), \quad (4)$$

where $z_0$ is the equilibrium interlayer separation for the commensurate AB stacking, $\Gamma_0$ is the adhesion energy, $a$ is the equilibrium carbon-carbon bond length in the graphene lattice, $G_1 = \frac{4\pi}{3}$, $\eta$ and $\beta$ are two dimensionless parameters. Here, the in-plane coordinates are set up such that the $x$-axis is parallel to a zigzag direction of the graphene lattice and the $y$-axis is parallel to an armchair direction (see Fig. 1a). As in Xue et. al (2022), we use the following values for the model parameters in numerical calculations: $a = 0.142$ nm, $z_0 = 0.334$ nm, $\Gamma_0 = 0.25$ J/m², $\eta = 0.0032$ and $\beta = 28.7$; these parameters were obtained previously based on atomistic calculations. We note that different forms of the interlayer potential energy function have been used by others for bilayer graphene and other 2D materials (Zhou et al., 2015; Dai et al., 2016a; Nam and Koshino, 2017; Carr et al., 2018).

Figure 1 (b-c) show the periodic corrugation of the interlayer potential energy along the armchair and zigzag directions, where the graphene lattice is assumed to be rigid and the interlayer separation is fixed with $u_z = 0$. With respect to the displacement in the armchair direction ($u_y$), the interlayer energy corrugation repeats with a period of $3a$, with a minimum for AB stacking ($u_x = u_y = 0$) and a maximum for AA stacking at $u_y = a$ (see Fig. 1b). The energy difference between the AB and AA stacking is about 0.1 J/m², although the interlayer potential energy for AA stacking can be reduced (from -0.15 to -0.184 J/m²) by slightly increasing the interlayer separation. In addition, there is a local maximum at $u_y = 2.5a$ for the saddle point (SP) stacking. As noted by Alden et al. (2013), the energy difference between the SP and AB stacking is around a factor of 10 lower than the energy difference between AA and AB stacking. With respect to the displacement in the zigzag direction ($u_x$), the interlayer energy is sinusoidal with a period of $\sqrt{3}a$ and a maximum lower than that for AA stacking but higher than for SP stacking (Fig. 1c).

We note that, while the potential function in Eq. (1) is sufficient to describe the interlayer energy corrugation with respect to the relative displacement between two rigid graphene layers, it ignores the effect of relative rotation between the two layers. In general, when twisting a graphene layer on top of another graphene layer, the relative motion at an arbitrary location consists of both translation and rotation. However, if the relative rotation is small (e.g., less than a few degrees), the effect is negligible. Thus, with the small-rotation assumption, we consider only the effect of translation by Eq. (1) in the present study for small-twist-angle graphene flakes.

## 3. Twisting of a rigid graphene flake

First consider a rigid graphene flake on top of a rigid graphene layer (Fig. 2a). Twisting the upper layer by an angle of rotation $\phi$ while holding the lower layer fixed, the relative in-plane displacements between the two layers are:

$$u_x(r,\theta) = r(\cos(\theta + \phi) - \cos(\theta)), \tag{5}$$
$$u_y(r,\theta) = r(\sin(\theta + \phi) - \sin(\theta)), \tag{6}$$

where $(r, \theta)$ are the polar coordinates with respect to the center of the flake. In addition, a relative normal displacement between the two rigid layers may be induced so that the interlayer separation changes to $z_0 + u_z$, where $u_z$ is to be determined as a function of the twist angle $\phi$. For the moment we assume that both graphene layers are rigid, whereas any deformation due to structural relaxation will be considered in Section 4.

Start with an infinitesimal twist angle (i.e., $r\phi \ll a$), so that the relative displacements are small everywhere in the graphene flake. In this case, we expand the interlayer potential function Eq. (1) into a Taylor series and retain the leading order terms only, yielding

$$U(u_x, u_y, u_z) \approx U_0(0) + \frac{1}{2}U_0''(0)u_z^2 + \frac{3}{8}U_1(0)\left(G_1\frac{r\phi}{a}\right)^2, \tag{7}$$

where $U_0''(0)$ is the second derivative of the function $U_0(u_z)$ at $u_z = 0$. The average interlayer potential energy per unit area of a circular graphene flake with a radius $R$ can be calculated as

$$\bar{U} = \frac{1}{A}\iint U dA \approx U_0(0) + \frac{1}{2}U_0''(0)u_z^2 + \frac{3}{16}U_1(0)\left(G_1\frac{R\phi}{a}\right)^2, \tag{8}$$

where $A = \pi R^2$.

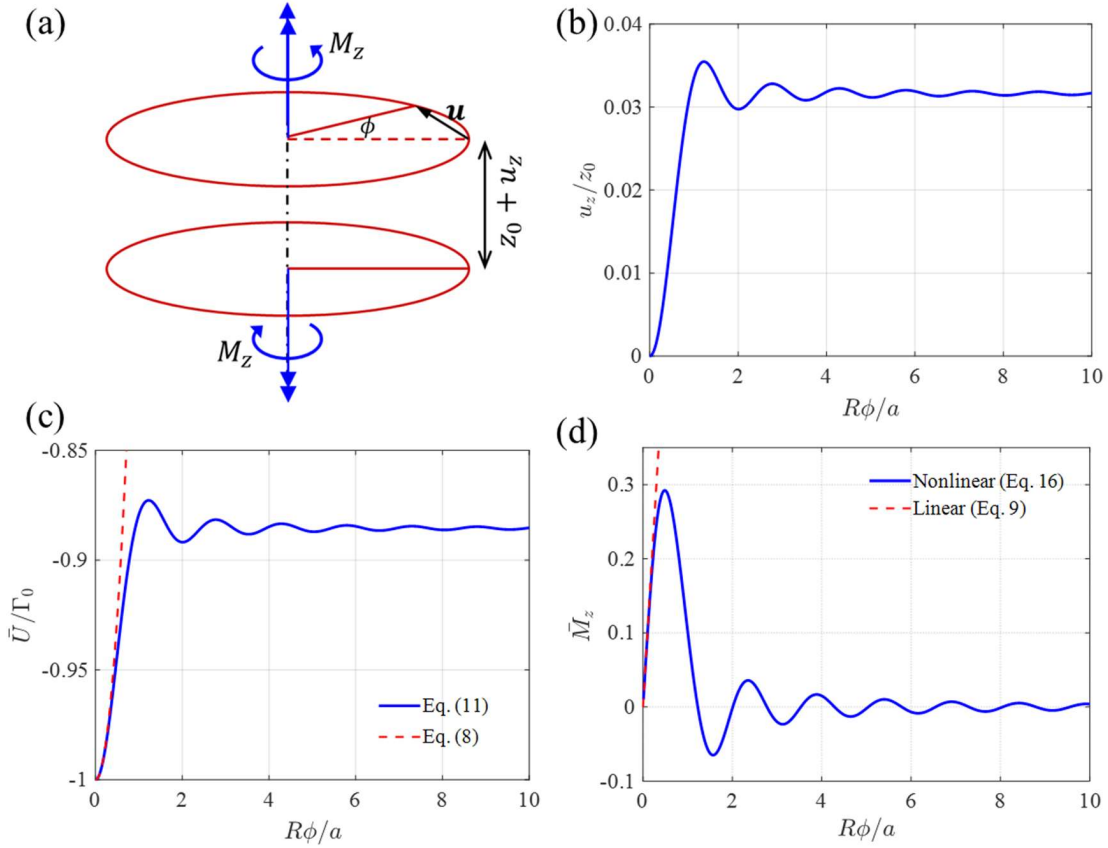

**Figure 2**. (a) Schematic of relative displacements between two rigid graphene layers subject to a twisting moment. (b) Change of the interlayer separation as a function of the twist angle. (c) Average interlayer potential energy (normalized by the adhesion energy) as a function of the twist angle. (d) Normalized twisting moment as a function of the twist angle.

Given a twist angle $\phi$, the average interlayer potential energy in Eq. (8) is minimized for $u_z = 0$, indicating no normal displacement for the case of an infinitesimal twist angle ($R\phi \ll a$). Taking the derivative of the average potential energy with respect to the twist angle, we obtain the twisting moment applied to the graphene flake as

$$M_z = A\frac{d\bar{U}}{d\phi} = \frac{2\pi^3}{3a^2}R^4 U_1(0)\phi. \qquad (9)$$

The linear relation between the twisting moment and the angle of rotation defines a torsional rigidity of the graphene bilayer that is proportional to the polar moment of the circular area, $I_p = \frac{1}{2}\pi R^4$. Following the classical mechanics of pure torsion, an interlayer shear modulus can be determined as

$$\mu = \frac{M_z z_0}{\phi I_p} = \frac{4\pi^2 z_0}{3a^2} U_1(0), \qquad (10)$$

which yields a value of 4.9 GPa by the present model, in close agreement with the values reported previously for single-crystal graphite from experiments (Nicklow et al., 1972; Bosak et al., 2007; Tan et al., 2012) and atomistic calculations (Mounet and Marzari, 2005; Michel and Verberck, 2008; Savini et al., 2011; Ruiz et al., 2015). Eq. (10) indicates that the interlayer shear modulus is proportional to the amplitude of energy corrugation, $U_1(0) = (\beta - 1)\eta\Gamma_0$. Thus, a measurement of the interlayer shear modulus (by twisting or sliding) could be used to determine the energy corrugation amplitude. In experiments, a twisting moment may be applied by an atomic force microscopy (AFM) tip to rotate the graphene flake (Ribeiro-Palau et al., 2018). The interlayer shear modulus may also be measured by inelastic x-ray scattering (Bosak et al., 2007) or Raman spectroscopy (Tan et al., 2012).

Even with a small twist angle, however, the relative in-plane displacements as given in Eqs. (5-6) can be large (relative to $a$) in a large flake so that the corrugation of the interlayer potential energy in Eq. (1) must be fully accounted for. In this case, we calculate the average interlayer potential energy of a circular graphene flake as

$$\bar{U} = \frac{1}{A}\iint U(u_x, u_y, u_z)dA = U_0(u_z) + U_1(u_z)\bar{F}(R, \phi), \tag{11}$$

with

$$\bar{F}(R, \phi) = \frac{3}{2} - \frac{3a}{2G_1 R \sin\left(\frac{\phi}{2}\right)} J_1\left(2G_1 \frac{R}{a}\sin\left(\frac{\phi}{2}\right)\right) \approx \frac{3}{2} - \frac{9a}{4\pi R\phi} J_1\left(\frac{4\pi}{3}\frac{R\phi}{a}\right), \tag{12}$$

where $J_\alpha(\cdot)$ is the Bessel function of the first kind. Note that $\bar{F}(R, \phi) \approx \frac{1}{3}\left(\frac{\pi R\phi}{a}\right)^2$ when $R\phi \ll a$, which recovers Eq. (8) for an infinitesimal twist angle. With the assumption of a small twist angle but relatively large in-plane displacements, the function $\bar{F}(R, \phi)$ can be written approximately as a function of a single variable, $R\phi/a$. Alternatively, with the moiré length scale (Shallcross et al., 2010; Hermann, 2012), $\lambda_M = \frac{\sqrt{3}a}{2\sin(\phi/2)}$, we may re-write $\bar{F}$ as a function of the ratio $\bar{R} = R/\lambda_M$, namely

$$\bar{F}(\bar{R}) = \frac{3}{2} - \frac{3\sqrt{3}}{4\pi\bar{R}} J_1\left(\frac{4\pi}{\sqrt{3}}\bar{R}\right). \tag{13}$$

For a small twist angle, we have $R\phi/a \approx \sqrt{3}\bar{R}$.

Given a twist angle $\phi$, the average interlayer potential energy in Eq. (11) is minimized by setting $\frac{\partial \bar{U}}{\partial u_z} = 0$, which predicts a normal displacement as

$$u_z = z_0\left[\left(\frac{2.5\beta\alpha+1}{\alpha+1}\right)^{\frac{1}{6}} - 1\right], \tag{14}$$

where $\alpha = \frac{3}{5}\eta\bar{F}(R\phi/a)$. The non-zero relative normal displacement leads to a change in the interlayer separation as a function of the twist angle (Fig. 2b). Such a dimensional change in the normal direction due to twisting is known as the Poynting effect commonly observed in finite deformation of elastomeric materials (Treloar, 1975). It is noted that the leading order of the normal displacement is quadratic in terms of $R\phi/a$, which vanishes in the linear analysis for an infinitesimal twist angle ($R\phi \ll a$). For relatively large values of $R\phi/a$, the interlayer separation approaches a constant of $1.032\, z_0 = 0.344\, nm$, about 3% larger than the equilibrium separation for the commensurate AB stacking of a graphene bilayer. Such a larger interlayer separation is expected for incommensurate stacking after twist. For comparison, the equilibrium separation is found to be 0.362 nm for AA stacking and 0.338 nm for SP stacking by the present model, both in good agreement with previously reported values (Jain et al., 2017; Uchida et al., 2014).

With the normal displacement ($u_z$) determined in Eq. (14) as a function of $R\phi/a$, the average interlayer potential energy ($\bar{U}$) in Eq. (11) is obtained also as a function of $R\phi/a$ for small twist angles, as shown in Fig. 2c. For $R\phi \ll a$, the interlayer energy density is quadratic with $R\phi/a$, as predicted by Eq. (8). For relatively large values of $R\phi/a$, the interlayer energy density changes non-monotonically, with local maxima and minima. In particular, the first three local minima of the interlayer potential energy correspond to $\frac{R\phi}{a} = 2.04, 3.54,$ and $5.06$, whereas the corresponding twist angles depend on the graphene flake size ($R$). It is also interesting to note that the maximum interlayer potential energy is around $-0.87\Gamma_0$ (~13% higher than the minimum at the commensurate AB stacking), which is reached at a twist angle, $\frac{R\phi}{a} = 1.23$. This angle may be considered as a critical twist angle. When the twist angle is smaller than the critical angle, the graphene flake would rotate back to the commensurate AB stacking ($\phi = 0$) once the applied moment is released. When the twist angle is greater than the critical angle, however, the graphene flake would remain twisted as the interlayer potential energy approaches a local minimum after the applied moment is released. The critical twist angle is thus inversely proportional to the flake size ($R$). A similar behavior was observed for twisted bilayer graphene and MoS$_2$ in MD simulations (Zhu et al., 2019 and 2021). For relatively large values of $\frac{R\phi}{a}$, the interlayer potential energy approaches a constant, which is around $-0.885\Gamma_0$. As expected, the incommensurate stacking (due to twist) increases the interlayer potential energy and hence reduces the adhesion energy compared to the commensurate AB stacking.

By taking derivative of the total interlayer potential energy with respect to the twist angle, we obtain the twisting moment (or torque) as

$$M_z = A\frac{d\overline{U}}{d\phi} = U_1(u_z)\left(\frac{3\pi R^2}{2\phi}\right)\left[\frac{2a}{G_1 R\phi}J_1\left(\frac{G_1 R\phi}{a}\right) - J_0\left(\frac{G_1 \phi R}{a}\right) + J_2\left(\frac{G_1 \phi R}{a}\right)\right]. \quad (15)$$

Normalizing the twisting moment by $\mu I_p a/(z_0 R)$, we obtain

$$\overline{M}_z = \frac{M_z z_0 R}{\mu I_p a} = \frac{U_1(u_z)}{\Gamma_0 \eta(\beta-1)}\left(\frac{9a}{4\pi^2 \phi R}\right)\left[\frac{2a}{G_1 R\phi}J_1\left(\frac{G_1 R\phi}{a}\right) - J_0\left(\frac{G_1 \phi R}{a}\right) + J_2\left(\frac{G_1 \phi R}{a}\right)\right], \quad (16)$$

which is a function of $\frac{R\phi}{a}$. As shown in Fig. 2d, the twisting moment is linear with respect to the twist angle for $\frac{R\phi}{a} \ll 1$, as predicted by Eq. (9). For relatively large values of $\frac{R\phi}{a}$, the twisting moment depends on the twist angle nonlinearly and non-monotonically. A maximum moment is predicted, after which the twisting moment decays and oscillates around zero. Corresponding to the local maxima and minima in the interlayer potential energy (Fig. 2c), the twisting moment becomes zero, meaning that the twisted graphene flake is in mechanical equilibrium with zero moment applied. However, only those equilibrium states corresponding to the local energy minima are stable (against small perturbations). Again, the corresponding twist angles for the stable equilibrium states depend on the graphene flake size ($R$). The maximum twisting moment is found to be: $M_{max} = 0.29\frac{\mu I_p a}{z_0 R}$, which is linearly proportional to the interlayer shear modulus and scales with the flake size as: $M_{max} \sim \mu R^3$. The maximum twisting moment may be considered as a critical moment (or twisting strength), corresponding to a twist angle at $\frac{R\phi}{a} = 0.49$. The presence of a maximum interlayer potential energy (Fig. 2c) and a maximum twisting moment (Fig. 2d) suggests that the transition from the commensurate AB stacking to an incommensurate twisted state or vice versa has a finite energy barrier and requires external stimuli, which may result in size-dependent thermal stability of the twisted flakes (Feng et al., 2013; Bagchi et al., 2020). Moreover, both the maximum energy and maximum moment would change considerably due to structural relaxation in the twisted graphene flake, as discussed in the next section.

## 4. Twisting of an elastic graphene flake

A graphene monolayer is highly flexible and thus the twisted graphene flake is expected to deform. Due to the interlayer coupling with inhomogeneous vdW interactions, the deformation of a twisted graphene flake is generally inhomogeneous, with both in-plane and out-of-plane components. Consequently, the graphene flake is no longer flat as assumed for a rigid flake, and the strain of the graphene lattice is inhomogeneous. To determine the inhomogeneous deformation of a twisted graphene flake, we employ a finite element method using the commercial software ABAQUS, similar to a previous study on sliding of graphene nanoribbons (Xue et al., 2022). The circular graphene flake is modeled by linearly elastic shell elements (S4R, element size ~ 0.1 nm), with an effective in-plane Young modulus $Et = 345$ N/m, Poisson's ratio $\nu = 0.16$, and a bending modulus $D = 1.5$ eV. In this study, we assume that the bottom graphene monolayer is attached to a rigid substrate so that it does not deform. The vdW interactions between the graphene flake and the bottom layer are simulated by the normal and shear tractions derived from the potential energy function in Eq. (1), which has been implemented as a user-defined subroutine (UINTER) in ABAQUS (Xue et al., 2022). Start with an equilibrium configuration where a circular graphene flake of radius $R$ is perfectly flat on top of a large graphene monolayer, with an interlayer separation of $z_0 = 0.334$ nm for the commensurate AB stacking. To twist the graphene flake, a circumferential displacement ($u_\theta = R\phi$) is imposed along its edge. At each twist angle, the elastic deformation of the graphene flake is determined by solving the boundary value problem. Due to the highly nonlinear and non-monotonic variation of the potential energy with respect to the displacements (see Fig. 1 b-c), the problem is numerically challenging. To mitigate numerical stability issues, we conduct nonlinear dynamic implicit simulations with a mass scaling method similar to the previous study (Xue et al., 2022). To minimize the rate effect, we apply sufficiently low rates for the twisting displacement so that the results are quasistatic with negligible kinetic energy (see Fig. A1 in Appendix).

The interlayer potential energy and the elastic strain energy of a twisted graphene flake ($R = 50$ nm) are calculated as functions of the twist angle shown in Fig. 3a. Compared to twisting a rigid flake of the same size, the interlayer potential energy is much lower for the elastic flake, but the elastic strain energy is higher. The total energy including both the interlayer and intralayer strain energy is lower than that for a rigid flake. Thus, by the elastic deformation of the graphene flake, the interlayer potential energy is reduced and the total energy is relaxed. Such a structural relaxation is thermodynamically favored and mechanically required to satisfy the local equilibrium everywhere in the elastic graphene flake. Interestingly, similar to the rigid flake, the total energy varies with the twist angle non-monotonically, with local minima and maxima corresponding to stable and unstable equilibrium states, respectively. Such an energy variation is similar to those obtained by atomistic calculations using a climbing-image nudged elastic band method for untwisting graphene flakes from a relatively large twist angle (Bagchi et al., 2020). Moreover, the elastic strain energy first increases and reaches its maximum at twist angles in the range of 0.2-0.5 degrees, after which the elastic strain energy decays and approaches zero at relatively large twist angles. This trend suggests that structural relaxation by elastic deformation is most significant at relatively small twist angles. For relatively large twist angles, the interlayer coupling becomes too weak to deform the elastic flake, so that the flake behaves like a rigid flake with negligible deformation. The similar trend was noted recently for twisted $MoS_2$ bilayers (Quan et al., 2021).

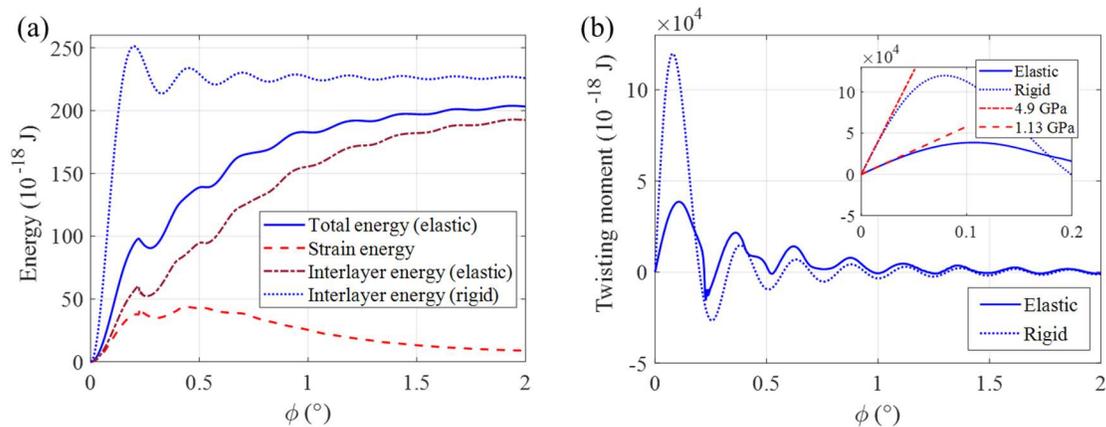

**Figure 3.** (a) The interlayer potential energy and the elastic strain energy of a circular graphene flake ($R = 50$ nm) versus the twist angle. The total energy includes both and is lower than the interlayer potential energy for a twisted rigid flake of the same radius. (b) Twisting moment as a function of the twist angle. Inset shows the linear regime for infinisimal twist angles with different values of the interlayer shear modulus for the rigid and elastic flakes.

The derivative of the total energy with respect to the twist angle yields the twisting moment applied along the edge of the circular flake, shown as a function of the twist angle in Fig. 3b. Compared to twisting a rigid graphene flake of the same size, the twisting moment is much lower for the elastic flake. Remarkably, even for an infinitesimal twist angle ($\phi < 0.1°$), the elastic deformation of the flake reduces the twisting moment considerably, yielding a lower torsional rigidity in the linear regime (see inset of Fig. 3b). The effective interlayer shear modulus deduced from the torsional rigidity by the classical relation ($\frac{M_z}{\phi} = \mu I_p/z_0$) is about 1.13 GPa, much lower than that predicted by Eq. (10) for a rigid flake. It is found that, unlike the rigid flake, twisting the elastic flake by a very small angle (e.g., $\phi = 0.05°$) leads to an inhomogeneous deformation of the elastic flake, where the center region is locked in the AB stacking with no rotation and the interlayer shear occurs only in an annular region near the edge of the flake (see Fig. A2 in Appendix). As a result, the classical assumption of the torsional deformation with a constant rotation of the entire flake is inapplicable for the elastic graphene flake. The initially localized deformation near the edge is similar to that for pulling a graphene nanoribbon (Xue et al., 2022), with an effective stiffness of the linear response depending on the flake size (Fig. A3). As the flake size decreases, the response approaches that of a rigid flake, while the effective interlayer shear modulus decreases as the flake size increases. Interestingly, an early experimental study (Soule and Nezbeda, 1968) reported interlayer shear modulus of single-crystal graphite ranging from 0.13 to 1.4 GPa and attributed the relatively low values to basal-plane dislocations. Moreover, compared to the rigid flake, the maximum twisting moment is also much lower for the elastic graphene flake (Fig. 3b). The first three twist angles corresponding to the stable equilibrium states (with zero twisting moment) are slightly different from those predicted for a rigid flake. The effect of elastic deformation diminishes as the twist angle increases (e.g., $\phi > 2°$).

To further understand the elastic deformation and structural relaxation of the twisted graphene flake, we show in Fig. 4 a sequence of contours for the evolving distributions of the interlayer potential energy and the elastic strain energy as the twist angle increases. The interlayer energy contours reflect the local stacking order, with the minimum energy (blue) for AB stacking and the maximum energy (yellow) for AA stacking, whereas the regions of SP stacking with an intermediate energy connects the AA domains and separates neighboring AB domains. The elastic strain energy contours show that the elastic deformation is localized in the SP domains with nearly zero strain energy in both the AB and AA domains. It is found that the strain state in the SP domains (or domain walls between AB domains) is nearly pure shear, and thus the SP domains form a network of shear strain solitons, as noted previously (Alden et al., 2013). At the first stable twist angle ($\phi = 0.274°$), three AA domains emerge as the vertices of an equilateral triangle moiré pattern at the center region of the graphene flake (Fig. 4 a-b). This is the largest stable moiré pattern that can form in the circular graphene flake with a radius of 50 nm, with the moiré length $\lambda_M \approx R$ ($\lambda_M = \frac{\sqrt{3}a}{2\sin(\phi/2)} = 51.4$ nm). At this twist angle, both the interlayer potential energy and the elastic strain energy drop from their maxima (Fig. 3a), yielding a local minimum for the total energy. Due to structural relaxation, a significant part of the graphene flake stays in the AB stacking, thus lowering the interlayer potential energy compared to the case of a rigid flake with no structural relaxation.

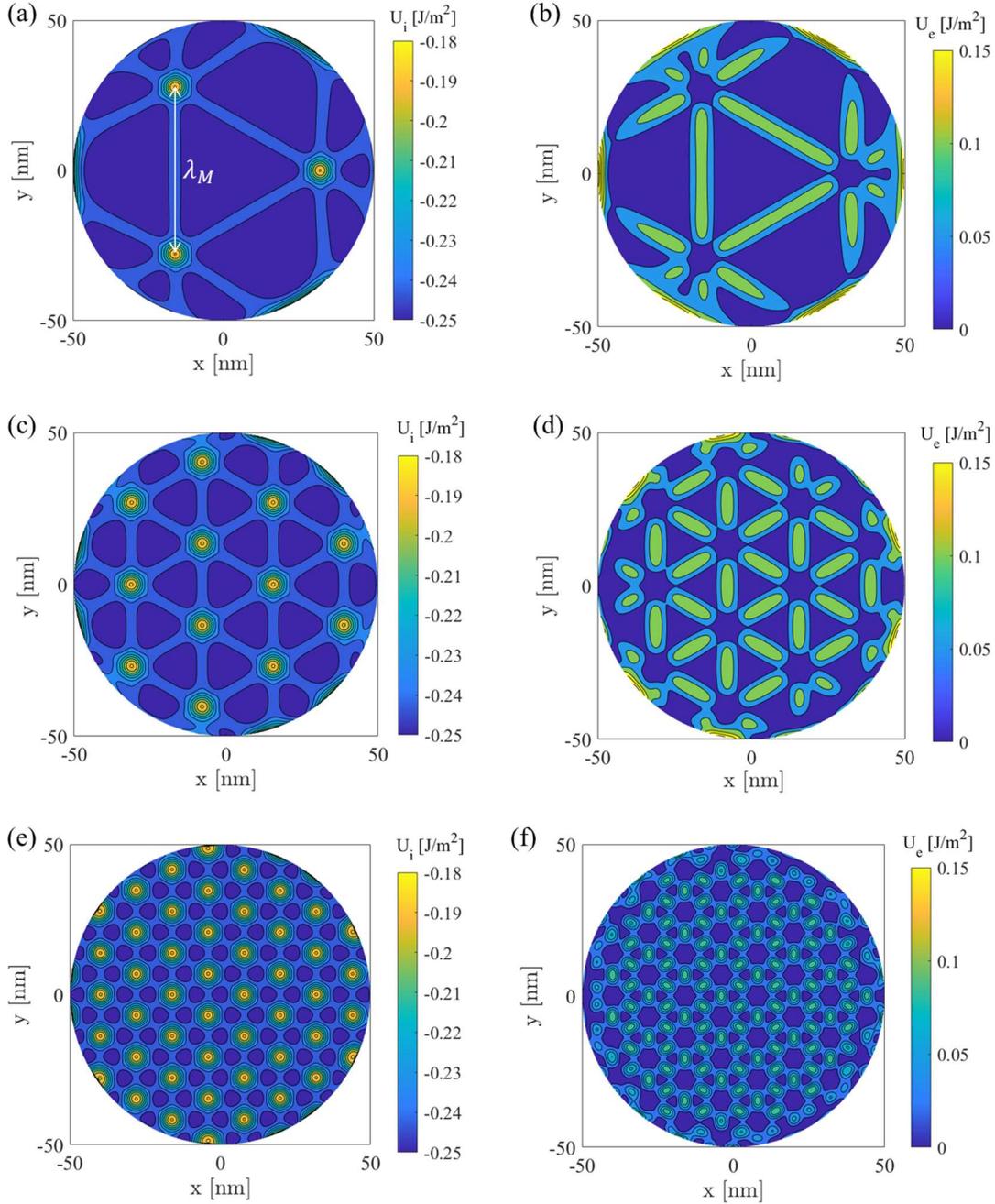

**Figure 4**. Contours of the interlayer potential energy (left) and elastic strain eneregy (right) of a twisted graphene flake ($R = 50$ nm). (a-b) At the first stable twist angle ($\phi = 0.274°$); (c-d) at the second stable angle ($\phi = 0.534°$); (e-f) at $\phi = 1.02°$.

In addition, the contours for the out-of-plane displacement ($u_z$) and the effective shear strain ($\gamma_{eff} = \sqrt{2e_{ij}e_{ij}}$, where $e_{ij}$ is the deviatoric strain components in 2D) are shown in Appendix (Fig. A4). The out-of-plane displacement is nearly zero ($u_z \approx 0$) in the AB domains and is the maximum ($u_z = 0.0277$ nm) in the AA domains, consistent with the different equilibrium separations for the two commensurate stacking orders. While the out-of-plane displacement is generally small, its effect on the interlayer coupling may not be negligible. As noted earlier, for AA stacking the interlayer energy is reduced by over 20% (from -0.15 to -0.184 J/m$^2$) due to the out-of-plane displacement. The effective shear strain is localized along the boundaries between the AB domains. The variation of the effective shear strain across a domain wall (strain soliton) is shown in Fig. A5 (b), with a maximum of 0.8% and a FWHM (full width at half maximum) of 10 nm for the twist angle $\phi = 0.274°$.

At the second stable twist angle ($\phi = 0.534°$), more AA domains emerge to form a moiré pattern with a smaller length scale ($\lambda_M = \frac{\sqrt{3}a}{2\sin(\phi/2)} = 26.4$ nm) as shown in Fig. 4 (c-d). As a result, the total interlayer potential energy is higher than Fig. 4a, while the total strain energy is similar (Fig. 3a). The maximum shear strain in the domain boundaries is higher (~1.3%), but the FWHM is narrower (~7.5 nm) as shown in Fig. A5 (d). As the twist angle increases further, the moiré length scale continues to decrease, whereas the interlayer potential energy increases and the elastic strain energy decreases. At $\phi = 1.02°$, a dense array of AA domains has formed with $\lambda_M = 13.8$ nm. This angle is close to the magic angle for twisted bilayer graphene (Guinea and Walet, 2019; Carr et al., 2019), where a series of extraordinary phenomena have been discovered recently, including correlated insulating states, unconventional superconductivity, and emergent ferromagnetism with anomalous Hall effect (Sun and Hu, 2020). At this twist angle, the maximum shear strain in the domain boundaries is ~1.1% and the FWHM is ~5 nm as shown in Fig. A5 (f). The maximum shear strain decays as the twist angle increases further (Fig. A6).

Besides the in-plane strain, we can also calculate the local in-plane rotation from the displacement field as: $\phi_{local} = \tan^{-1}\left(\frac{\partial u_y}{\partial x} - \frac{\partial u_x}{\partial y}\right)$. In the case of a rigid flake, the local rotation is identical to the applied twist angle ($\phi_{local} = \phi$). For an elastic graphene flake, however, structural relaxation leads to inhomogeneous rotation, with a distribution of the locally differential angle of rotation, defined as the difference between the local and applied rotation angles, $\phi_d = \phi_{local} - \phi$, as shown in Fig. 5a for a circular graphene flake ($R = 50$ nm) at the second stable twist angle ($\phi = 0.534°$). Interestingly, the AB domains all have a negative differential rotation ($\phi_d < 0$) and nearly zero local rotation ($\phi_{local}^{AB} \approx 0$), whereas the AA domains all have a positive differential rotation ($\phi_d > 0$) and thus a local rotation greater than the applied twist angle ($\phi_{local}^{AA} > \phi$). In other words, compared to a constant rigid-body rotation, the AB domains rotate back and the AA domains rotate further, maintaining the same average rotation over the area of the graphene flake. As a result of the inhomogeneous local rotations, the direction of the in-plane displacement flips across the boundary between two neighboring AB domains, leading to a nearly pure shear deformation, as shown more clearly in Fig. 5b. The AA domains at the intersections of the shear strain solitons have been recognized as topological point defects, analogous to vortices in a superconductor (Alden et al., 2013). The local rotation angles ($\phi_{local}$) at the centers of the AB and AA domains depend on the applied twist angle as shown in Fig. 5c. The rotation of the AB domains remains nearly zero for small twist angles (up to ~0.8°), whereas the rotation of the AA domains remains nearly a constant (~1.36°) for twist angles from 0.27° to 0.8°. For smaller twist angles ($\phi < 0.27°$), the AA domains are not fully developed in the flake, and the partially developed AA domains are located near the edge of the flake with a smaller local rotation. For larger twist angles ($\phi > 0.8°$), the local rotation increases in both the AB and AA domains, and they converge toward the applied twist angle, approaching uniform rotation of a rigid flake. Therefore, the locally inhomogeneous rotation as a result of structural relaxation is most significant for small twist angles up to ~2°. We note qualitatively similar results in Zhang and Tadmor (2018) but two quantitative differences: (1) The relative rotation of the AB domains was larger for small twist angles (up to ~0.8°), and (2) the local rotation of the AA domains was also larger (~1.8°) for twist angles from 0.2° to 0.8°. These differences may be partly due to the finite flake size considered in the present work as opposed to the periodic boundary conditions in Zhang and Tadmor (2018).

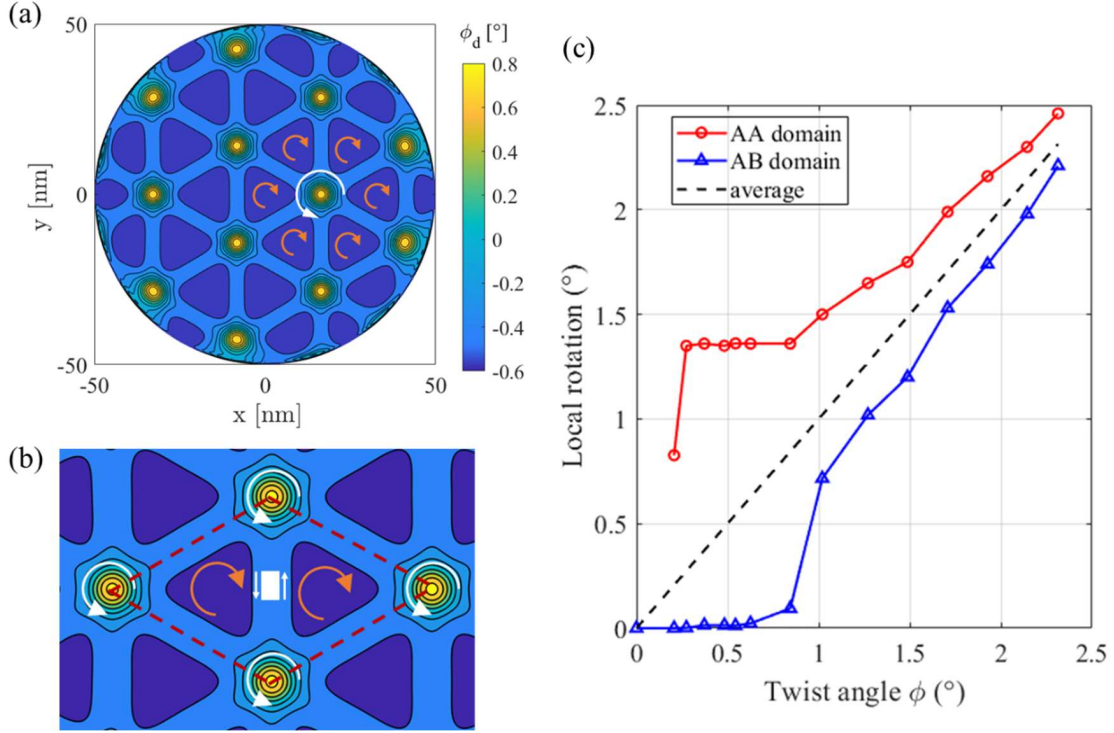

**Figure 5.** (a) Contour of locally differential angle of rotation in a circular graphene flake ($R = 50$ nm) at a twist angle ($\phi = 0.534°$), and (b) a close-up view of the local rotation in a moiré supercell. (c) Local rotation at the centers of the AA and AB domains versus the twist angle.

There are two fundamental length scales in the present model, the carbon-carbon bond length of the graphene lattice ($a = 0.142$ nm) and the equilibrium interlayer separation ($z_0 = 0.334$ nm), both atomistic scales. With a small twist angle, the moiré length scale, defined as the distance between neighboring AA domains as shown in Fig. 4a, is much larger ($\lambda_M = \frac{\sqrt{3}a}{2\sin(\phi/2)} \gg a$) and thus can be resolved by the continuum model of graphene and the finite element method. As shown in Fig. 6a, the moiré length scale obtained from the finite element model is in excellent agreement with the theoretical prediction. The moiré length is unaffected by the elastic lattice deformation and structural relaxation within each moiré supercell. For a finite graphene flake size ($R$), the largest stable moiré length is approximately equal to the flake radius (Fig. 4a). Another length scale of interest is the width of the strain soliton or domain wall between the neighboring AB domains. A previous study (Alden et al., 2013) found that the width of a shear strain soliton in bilayer graphene was around 6 nm based on atomic-resolution STEM measurements.

Theoretically, the width of a shear strain soliton was predicted by a two-chain Frenkel–Kontorova model (Popov et al., 2011; Alden et al., 2013) as

$$w_{sh} = \frac{a}{2}\sqrt{\frac{Et}{2(1+\nu)V_{sp}}} \quad (17)$$

where $V_{sp}$ is the saddle-point energy per unit area (relative to the AB stacking). In the present model, $V_{sp} = 0.011$ J/m² (~1.6 meV/atom), and by Eq. (17) we obtain $w_{sh} = 8.25$ nm. In our numerical simulations, the width of the shear strain soliton can be determined based on the profiles of the interlayer potential energy or the effective shear strain across the soliton between two neighboring AB domains. A few such profiles are shown in Appendix (Fig. A4) for three twist angles. Note that the width of the strain soliton determined by the FWHM of the interlayer potential energy profile is slightly smaller than that determined by the FWHM of the shear strain profile. Both decrease as the twist angle increases (Fig. 6b). In contrast, for a rigid graphene flake, the FWHM of the interlayer potential energy profile is proportional to the moiré length, $w_{sh,rigid} \sim 0.3\lambda_M$. Figure 6b shows that the width of strain soliton (by the interlayer energy profile) is considerably smaller after structural relaxation for small twist angles ($\phi < 1°$), whereas the effect of structural relaxation diminishes for larger twist angles. Similarly, the width of the AA domain can be determined from the FWHM of the interlayer potential energy profile, which is $\sim 0.54\lambda_M$ for a rigid flake but again much smaller after elastic relaxation for small twist angles ($\phi < 1°$) (see Fig. A7a). The normalized FWHM widths of the SP and AA domains shown in Figure A7b are similar to a plot (Fig. 8b) in Annevelink et al. (2020), but the AA domain width appears to differ by a factor of $\sqrt{3}$ at relatively large twist angles ($\phi > 1.5°$), approaching the analytical prediction ($\sim 0.54\lambda_M$) for a rigid flake. Therefore, the width of the strain soliton sets the minimum length scale that should be numerically resolved in the finite element model with sufficiently fine meshes.

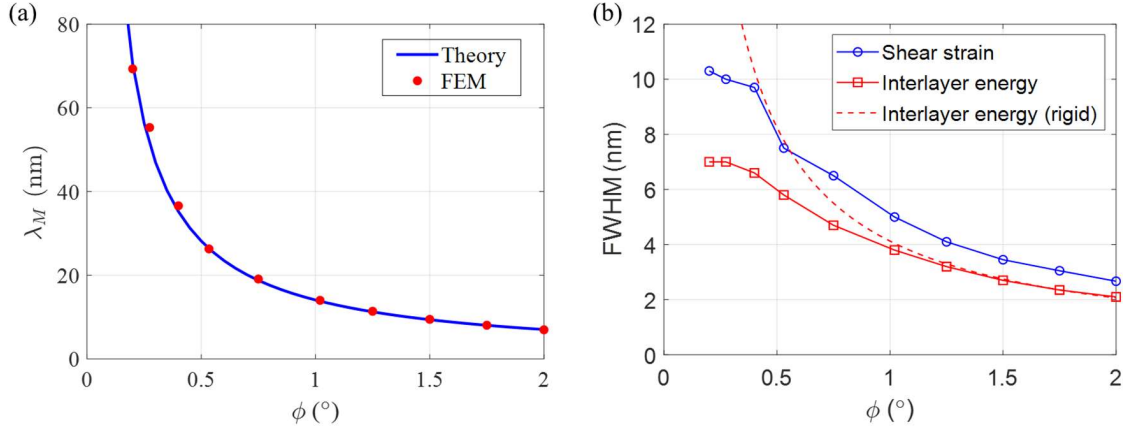

**Figure 6.** (a) Comparison of the moiré length scale in a twisted graphene flake ($R = 50$ nm) calculated from FE simulations and the theoretical prediction; (b) Width of strain soluton determined from the FWHM of the interlayer potential energy and the effetive shear strain profiles.

The finite size of the graphene flake considered in the present study introduces several size effects. First, there exists a critical twist angle, below which the graphene flake would rotate back to fully commensurate AB stacking once the twisting moment is released. The total energy reaches the first peak at the critical angle (Fig. 3a). For a rigid graphene flake, the critical angle is inversely proportional to the radius, $\phi_{critical} = 1.23a/R$ (Fig. 7a). For an elastic graphene flake, the critical angle is similar but slightly larger. Second, there exist a set of metastable twist angles, corresponding to the local minima in the total energy (Fig. 3a) and zero twisting moment (Fig. 3b). Again, for a rigid graphene flake, the metastable twist angles are inversely proportional to the radius (Fig. 7b), similar to the geometrical scaling relations obtained for triangular and hexagonal flakes by Zhu et al. (2021). For an elastic graphene flake, the metastable twist angles are slightly smaller. The difference results from the elastic deformation of the graphene lattice, which is most notable for small twist angles and relatively large graphene flakes (R > 10 nm). Moreover, passivation or reconstruction of the graphene edges (Acik and Chabal, 2011; Gan and Srolovitz, 2010) could introduce an additional size effect, which is, however, not considered in the present study.

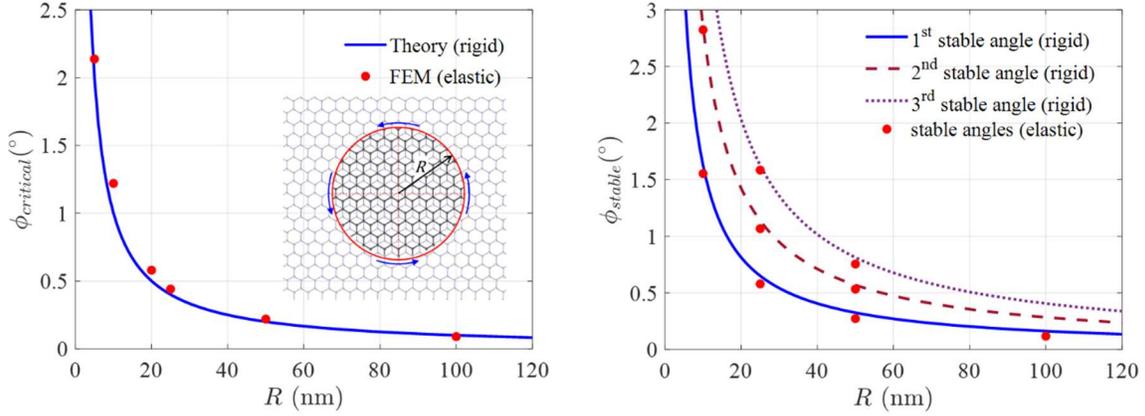

**Figure 7**: (a) The critical twist angle, and (b) the stable twist angles of circular graphene flakes, depending on the radius. The inset in (a) shows schematically a circular graphene flake on a graphene monolayer.

## 5. Summary

In this work, we present a continuum model for small-twist-angle graphene flakes on a rigid graphene substrate. The van der Waals interactions between the graphene layers are described by a periodic potential energy function. If the graphene flake is treated as rigid, the interlayer coupling leads to non-monotonic changes in the interlayer potential energy with a series of local minima. When the graphene flake is treated as an elastic membrane, the interlayer potential energy is reduced by inhomogeneous deformation of the graphene lattice with localized shear strain along the boundaries between neighboring domains of commensurate AB stacking. The structural relaxation and induced strain localization are most significant in a relatively large graphene flake at small twist angles. It is found that there are many metastable equilibrium configurations at particular twist angles, depending on the flake size.

**Acknowledgments**

VM, KML and RH gratefully acknowledge financial support by the Portuguese Foundation for Science and Technology – FCT under the UT Austin Portugal program through the project Soft4Sense. ZX acknowledges support from China Scholarship Council (CSC) for visiting the University of Texas at Austin. RH also acknowledges helpful discussions with Dr. Allan H. MacDonald from the Physics department of UT Austin.

**Appendx. Supplementary figures**

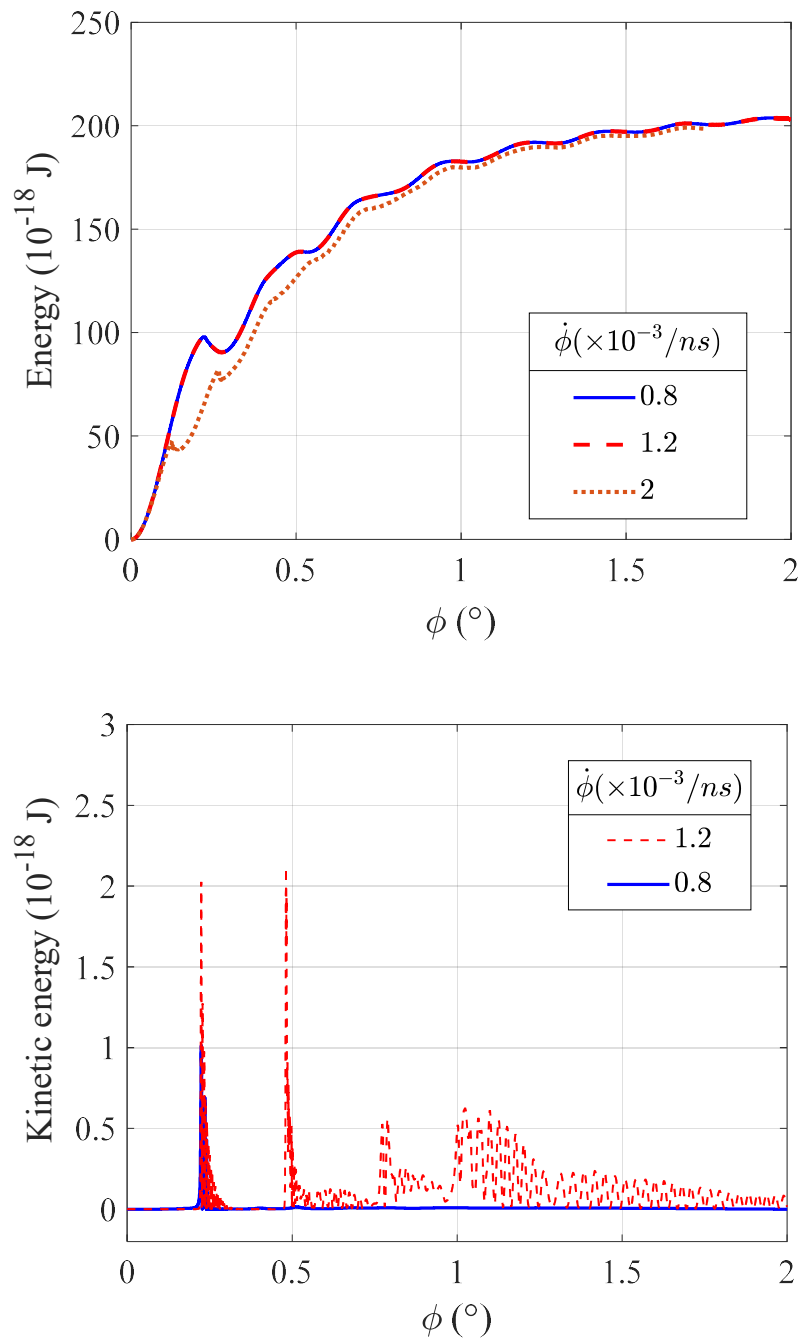

**Figure A1.** (a) The total energy (interlayer potential energy + elastic strain energy) in a circular graphene flake ($R$ = 50 nm) versus the twist angle, simulated at different rates of twisting. (b) The kinetic energy is much smaller for the two low rates of twisting, but becomes more significant at the higher rate (not shown).

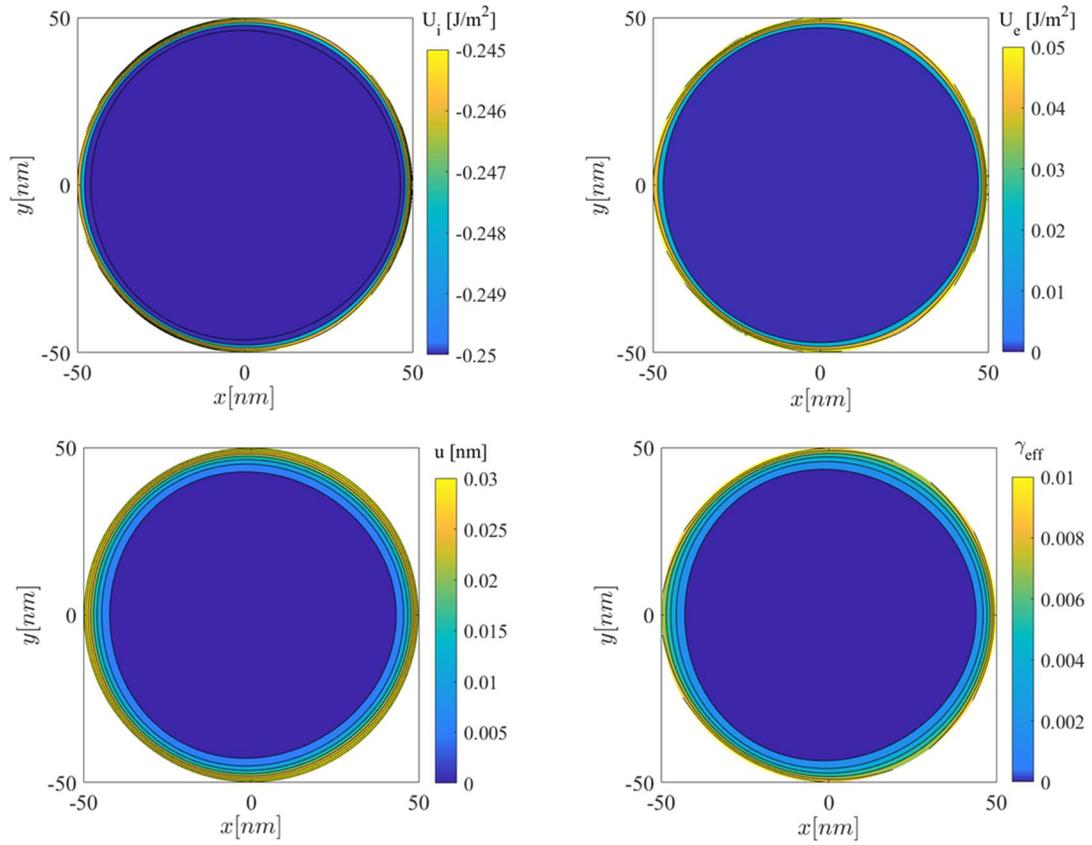

**Figure A2.** Contours at a small twist angle ($\phi = 0.05°$) showing interlayer shear and elastic deformation localized near the outer edge of a graphene flake ($R = 50$ nm), with no rotation or deformation at the center region. (a) Interlayer energy; (b) Elastic strain energy; (c) Displacement magnitude; (d) Effect shear strain.

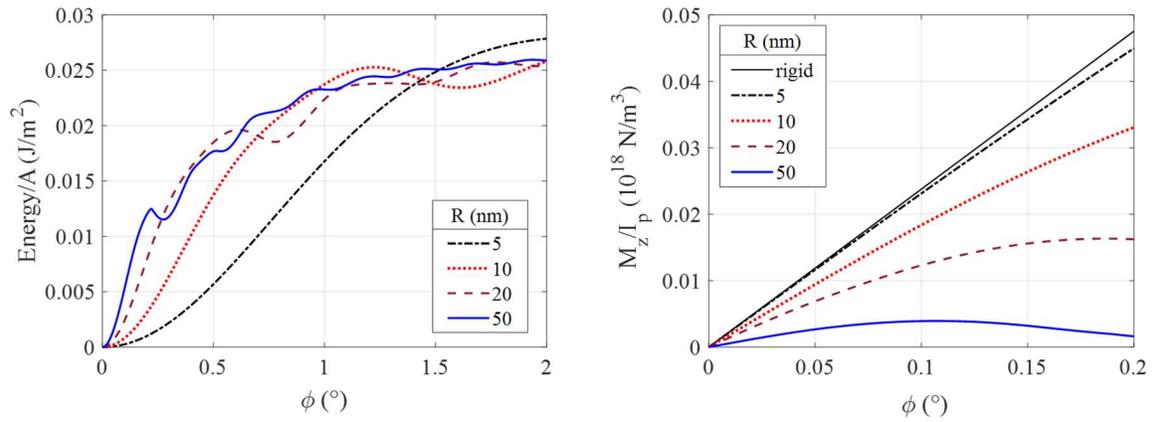

**Figure A3.** Effect of flake size: (a) Total energy (per unit area) as a function of the twist angle for circular graphene flakes of different radii. (b) Twisting moment as a function of the twist angle, showing the linear regime with size-dependent stiffness due to the localized deformation in the elastic flakes.

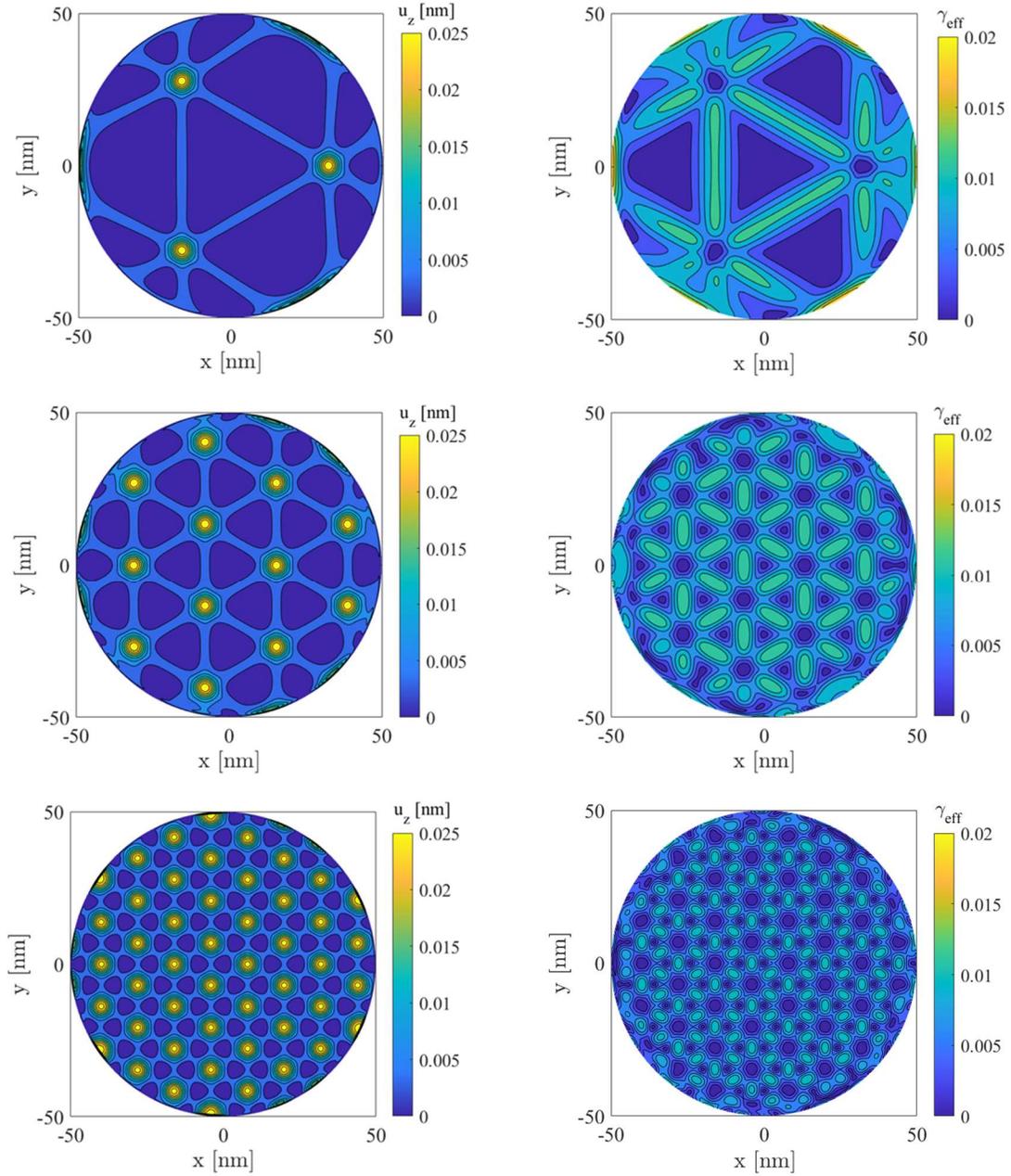

**Figure A4**. Contours of the out-of-plane displacement (left) and effective shear strain (right) of a twisted graphene flake ($R = 50$ nm). (a-b) At the first stable twist angle ($\phi = 0.274°$); (c-d) at the second stable angle ($\phi = 0.534°$); (e-f) at $\phi = 1.02°$.

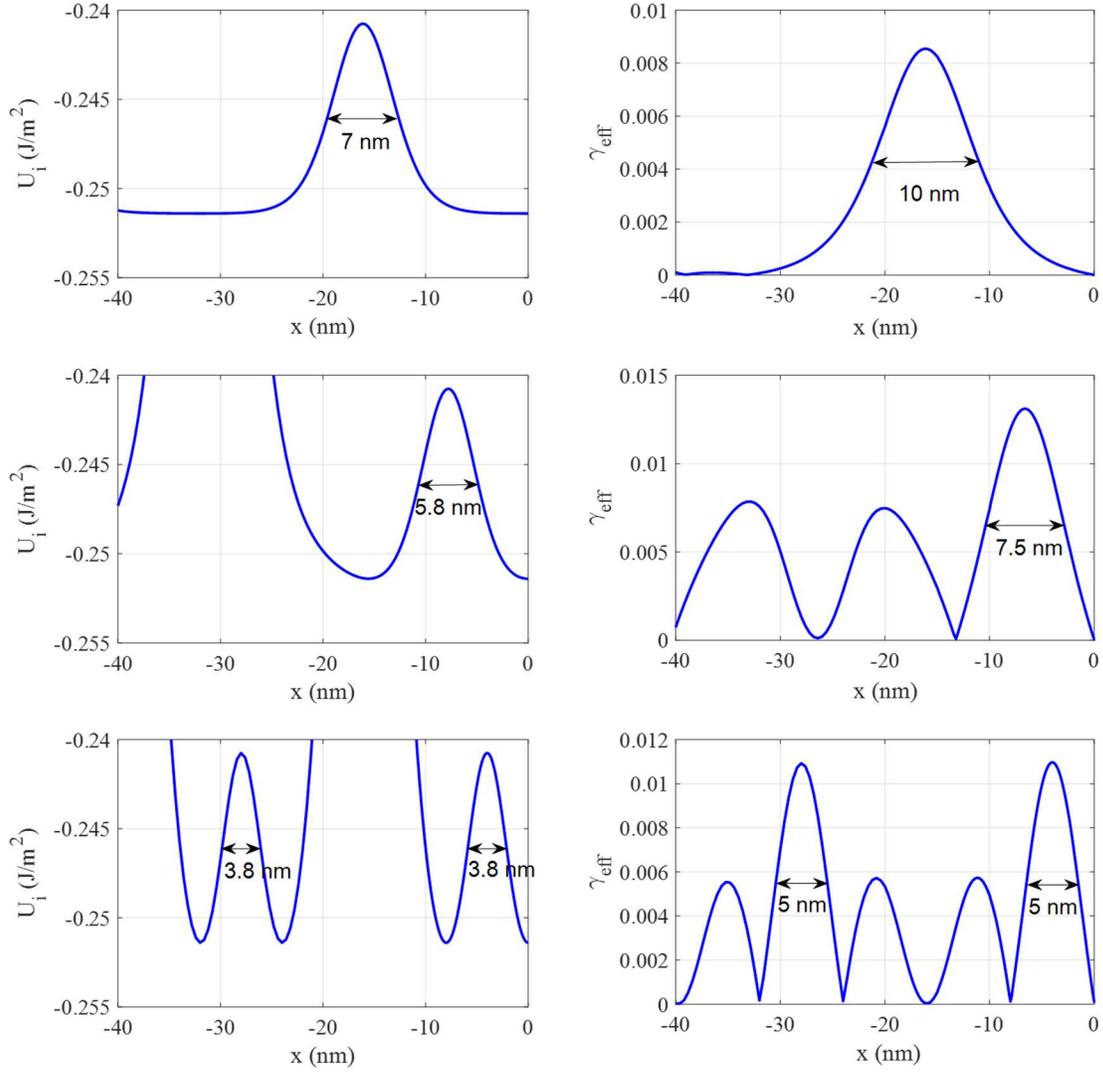

**Figure A5**. Profiles of the interlayer potential energy (left) and effective shear strain (right) along a horizontal line ($y = 0$) for a twisted graphene flake ($R = 50$ nm). (a-b) At the first stable twist angle ($\phi = 0.274°$); (c-d) at the second stable angle ($\phi = 0.534°$); (e-f) at $\phi = 1.02°$.

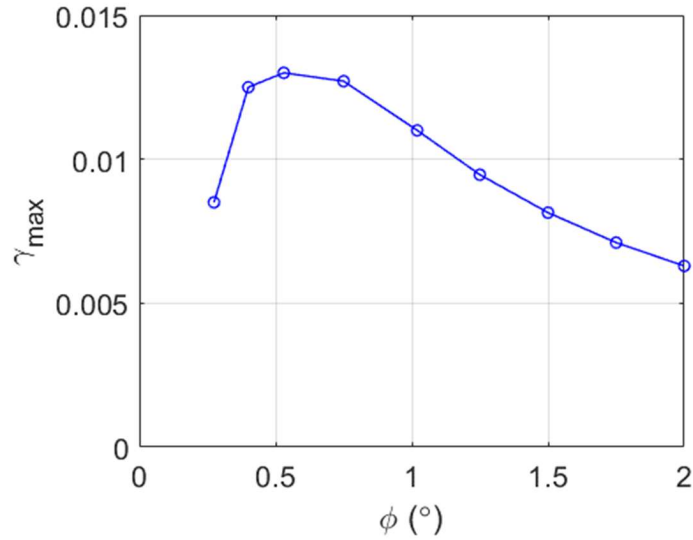

**Figure A6**. Maximum shear strain in the strain soliton (domain wall) between neighboring AB domains, calculated for a twisted graphene flake ($R = 50$ nm).

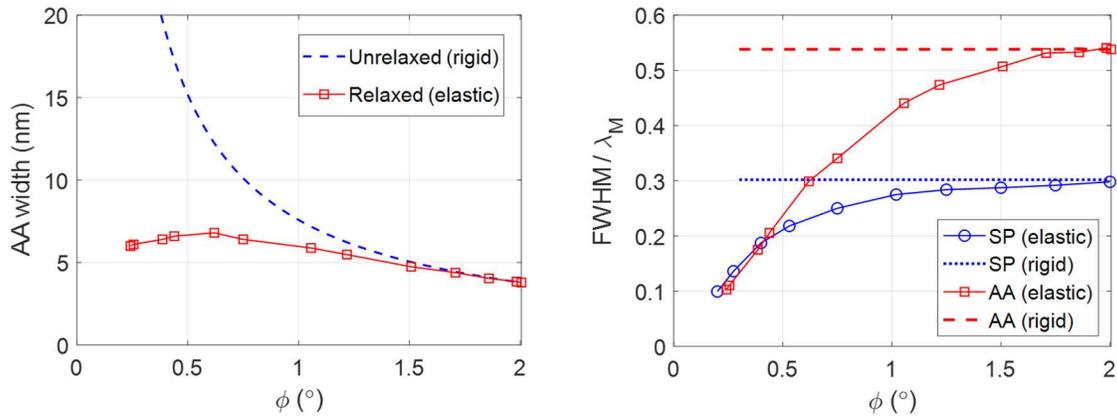

**Figure A7**. (a) FWHM width of the AA domain vs the twist angle; (b) Normalized FWHM width of the AA and SP domains. For a rigid graphene flake, the domain sizes can be calculated exactly based on the interlayer potential energy, which are proportional to the moiré length, $0.3\lambda_M$ for the SP domain (soliton) and $0.54\lambda_M$ for the AA domain. Both domain sizes are reduced by elastic deformation of the graphene flake, especially for small twist angles.